\documentclass[preprint,12pt]{elsarticle}
\biboptions{sort&compress}

\usepackage{amssymb}
\usepackage{amsmath}
\usepackage{bm}
\usepackage{physics}
\usepackage{etoolbox}
\usepackage{xcolor}

\newcommand{\del}[1]{}

\usepackage{hyperref}
\usepackage{doi}

\journal{Journal of Magnetism and Magnetic Materials}

\begin{document}

\begin{frontmatter}

\title{Theory of spin Seebeck effect activated by acoustic chiral phonons}
\author[inst1]{Naoki Nishimura} 
\author[inst2]{Takumi Funato}
\author[inst2,inst3,inst4,inst5]{Mamoru Matsuo}
\author[inst1]{Takeo Kato} 

\affiliation[inst1]{organization={Institute for Solid State Physics, The University of Tokyo},
addressline={5-1-5 Kashiwanoha}, 
city={Kashiwa},
postcode={277-8581}, 
state={Chiba},
country={Japan}}
\affiliation[inst2]{organization={Advanced Science Research Center, Japan Atomic Energy Agency},
addressline={2-4 Shirakata-Shirane, Tokai-mura}, 
city={Naka-gun},
postcode={319-1195}, 
state={Ibaraki},
country={Japan}}
\affiliation[inst3]{organization={Kavli Institute for Theoretical Sciences, University of Chinese Academy of Sciences},
addressline={No. 19 Yuquan Road, Haidian District}, 
city={Beijing},
postcode={100190}, 
country={China}}
\affiliation[inst4]{organization={CAS Center for Excellence in Topological Quantum Computation, University of Chinese Academy of Sciences},
addressline={No. 80 Zhongguancun East Road, Haidian District}, 
city={Beijing},
postcode={100190}, 
country={China}}
\affiliation[inst5]{organization={RIKEN Center for Emergent Matter Science (CEMS)},
addressline={2‑1 Hirosawa}, 
city={Wako},
postcode={100190}, 
state={Saitama},
country={Japan}}

\begin{abstract}
We theoretically explore the generation of spin current driven by a temperature gradient in a junction between a chiral insulator and a normal metal. Based on the gyromagnetic response induced by microscopic acoustic-phonon-mediated lattice rotation, we derive a formula for the spin current when a finite temperature difference is imposed between two ends of the sample. 
We clarify how the phonon-mediated spin current depends on the sample geometry, the thermal conductivity, the heat conductance at the interface, and the average temperature.
Our formulation provides a microscopic foundation for the chiral-phonon-activated spin Seebeck effect without relying on magnetism or spin-orbit interactions.
\end{abstract}


\end{frontmatter}

\section{Introduction}
\label{sec1}

Spin caloritronics~\cite{Bauer2012,Hoffmann2015,Uchida2021} explores the interaction between spin and heat currents, enabling the transformation of heat into spin currents, and vice versa, in various magnetic systems~\cite{Uchida2008}. This field of study offers a powerful approach to converting waste heat into spin currents, which could pave the way to the development of energy-efficient technologies.
One of the key phenomena in this field is the spin Seebeck effect~\cite{Uchida2008,Jaworski2010,Slachter2010,Uchida2011,Jaworski2012,Adachi2013}, a spin counterpart to the Seebeck effect in thermoelectricity, which, under a thermal gradient, leads to the appearance of non-equilibrium spin currents in an adjacent conductor.

\begin{figure}[tb]
\centering
\includegraphics[width=80mm] {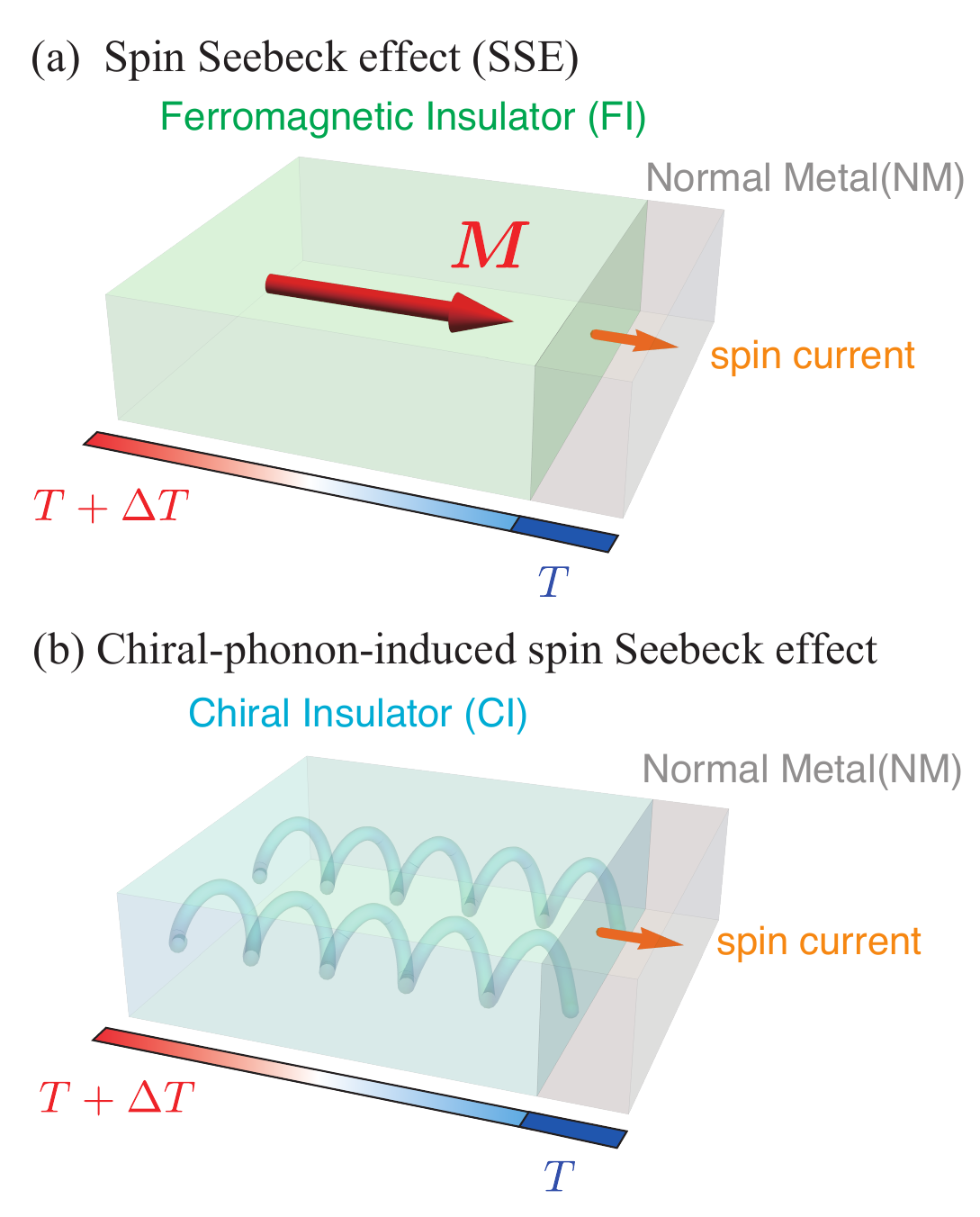}
\caption{\label{fig:schematic} (a) The conventional spin Seebeck effect in a junction composed of a ferromagnetic insulator (FI) and a normal metal (NM). 
(b) The chiral-phonon-activated spin Seebeck effect in a junction composed of a chiral insulator (CI) and a NM.
In both cases, the spin current is generated in the NM under a temperature bias between the two ends of the junction.}
\end{figure}

The spin Seebeck effect appears in magnetic junctions. Fig.~\ref{fig:schematic}(a) schematically shows spin Seebeck effect in a junction composed of a ferromagnetic insulator (FI) and a normal metal (NM), resulting to spin injection into the NM.
In a recent experiment~\cite{Kim2023}, however, it has been demonstrated that chiral phonons in a {\it non-magnetic} chiral insulator, which is an insulator with broken inversion symmetry, can generate spin currents into an adjacent {\it non-magnetic} metal when a thermal gradient exists (see Fig.~\ref{fig:schematic}(b)). 
In that experiment, the sign of the generated spin current changed depending on the chirality of the crystal structure~\cite{Barron2012,Kishine2022,Fransson2022}.
This phenomenon, named the `chiral-phonon-activated spin Seebeck effect', indicates the potential of using chiral phonons in spin-caloritronic applications and offers a new route towards spin generation~\cite{Zhang2025b,Zhang2025}. Compared with the spin Seebeck effect, the chiral-phonon-activated spin Seebeck effect has advantages for the development of spintronic devices with low environmental impact by the use of only light elements.
On the other hand, the extent to which spin currents can be enhanced remains largely unexplored.
To address this issue, it is crucial to elucidate the underlying microscopic mechanisms.

Chiral phonons in solids have been discussed mainly from the viewpoint of phonon angular momentum~\cite{Zhang-Niu2014}, whose foundational ideas can be traced back to earlier works~\cite{Vonsovskii1962}. 
The phonon angular momentum has been studied both theoretically and experimentally through several physical phenomena such as spin-phonon coupling via spin anisotropy~\cite{Garanin-Chudnovsky2015,Nakane-Kohno2018,Geilhufe2022}, ultrafast response by high-intensity THz lasers~\cite{Geilhufe2023,Basini2024,Davies2024,Choi2024}, magnetic effects on chiral phonons~\cite{Juraschek2019,Xiong2022a,Xiong2022b}, phonon transport across interfaces~\cite{Streib2018,Chen2022,Suzuki2024}, and angular momentum generation by the temperature gradient~\cite{Hamada2018,HZhang2025}.
Spin transport induced by the circular motion of atoms in the presence of chiral phonon modes has also been extensively explored in two-dimensional electron systems~\cite{Zhang-Niu2015,Chen2018,Ren2021,Saparov2022,Yao2025} and in three-dimensional chiral crystals~\cite{Yao2022,Yao2024,Kato2024}.
Moreover, the concept of pseudo-angular momentum~\cite{McLellan1988}, based on the discrete rotational symmetries preserved in crystal structures, has been employed to understand the selective excitation of chiral phonons~\cite{Bozovic1984,Tatsumi2018,Zhang2022,Komiyama2022,AKato2022,Tsunetsugu2023,AKato2023,Tateishi2025}, and the selective excitation has been confirmed experimentally~\cite{Zhu2018,Jeong2022,Ishito2023a,Ishito2023b}.
However, a key question has remained unanswered for a long time, i.e. whether a flow of chiral phonons under a thermal gradient can directly induce the spin Seebeck effect.

In this study, we propose a microscopic mechanism underlying the chiral-phonon-activated spin Seebeck effect that is based on interfacial phonon-spin conversion due to the gyromagnetic effect~\cite{Barnett1915,Einstein-deHaas1915,Scott1962}.
We derive a formula for the phonon-mediated spin current flowing at an interface and discuss its features.

\section{Formulation}

\begin{figure}[tb]
\centering
\includegraphics[width=75mm] {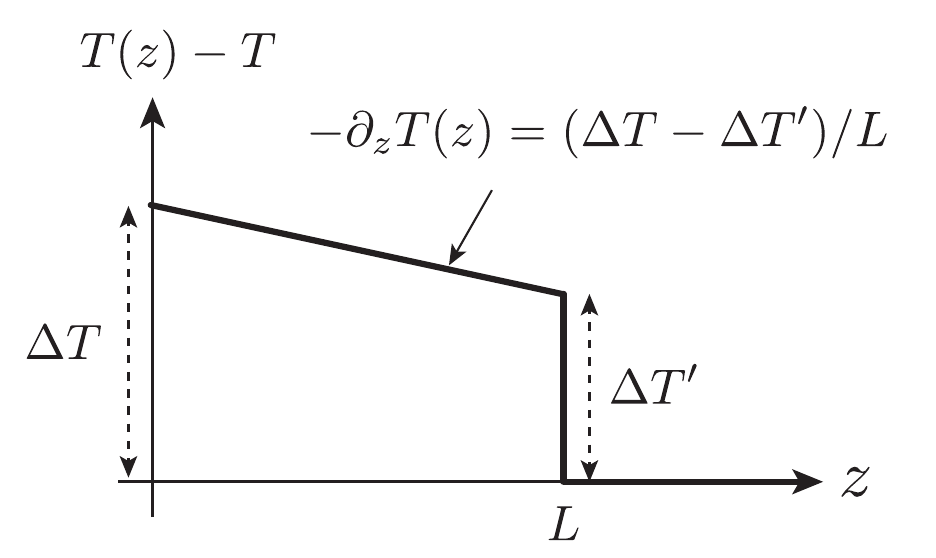}
\caption{\label{fig:temp} Temperature variation in the sample.}
\end{figure}

We consider a junction composed of a normal metal (NM) and a chiral insulator (CI), i.e., an insulator with a chiral crystal structure as shown in Fig.~\ref{fig:schematic}(b).
As illustrated in Fig.~\ref{fig:temp}, the $z$ axis is along the chiral axis of the CI, which is assumed to be perpendicular to the NM-CI junction interface, and the regions of the CI and NM are $0<z<L$ and $L<z$, respectively.
We impose a temperature difference $\Delta T$ between the ends of the sample, as shown in Fig.~\ref{fig:schematic}(b).
Furthermore, we assume that the temperature in the NM is constant due to the large thermal conductivity and that the spatial gradient of the temperature in the CI is constant, assuming a sufficiently thin CI layer.
Denoting the temperature drop at the interface as $\Delta T'$, the temperature gradient in the CI is given as $\partial_z T(z) = -(\Delta T-\Delta T')/L$.

\begin{figure}[tb]
\centering
\includegraphics[width=60mm] {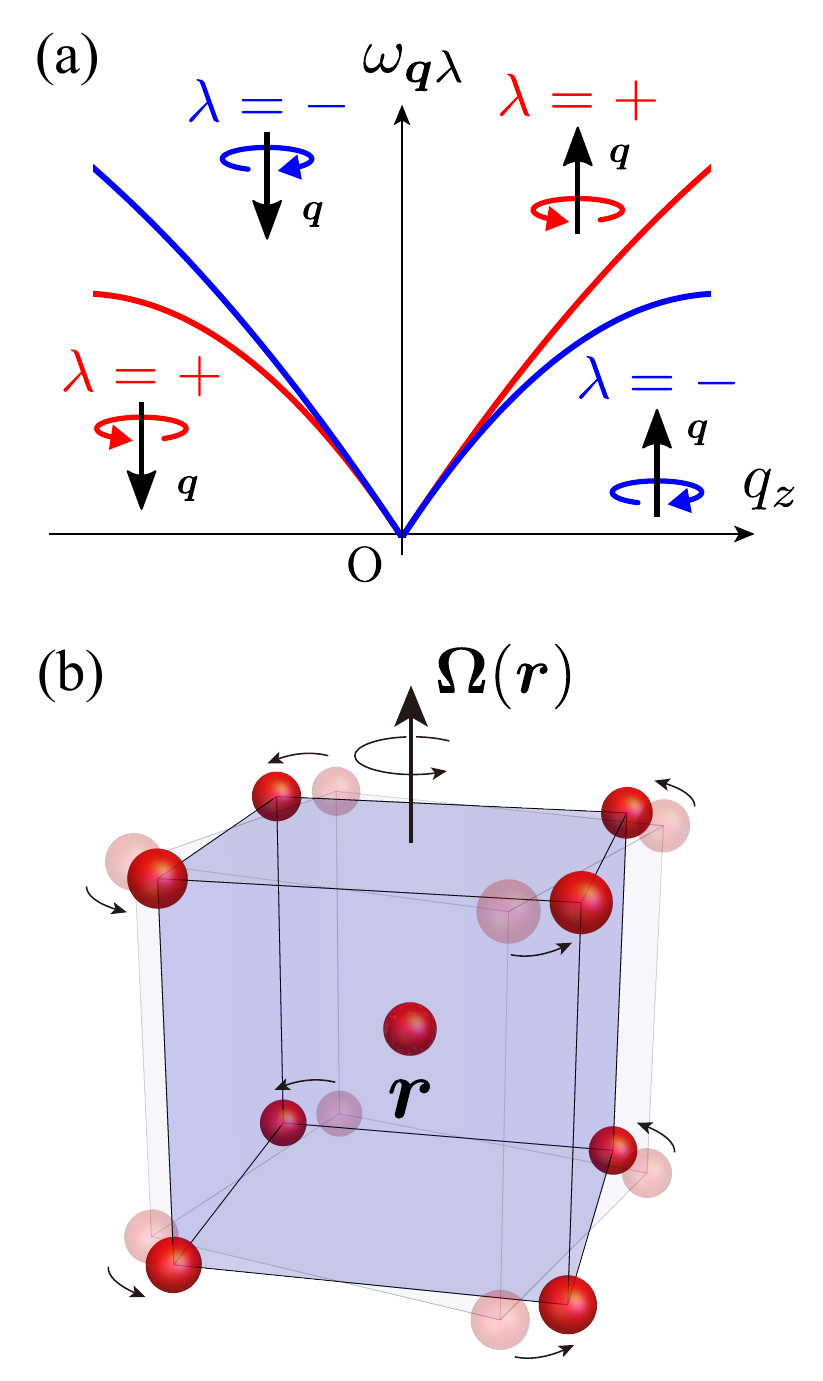}
\caption{\label{fig:phonon} (a) Schematic picture of dispersion of chiral phonons. (b) Schematic picture of microscopic rotation by chiral phonons. The vorticity $\Omega({\bm r})$ is related to the rotation axis and the angular velocity of the relative motion of the atoms around the atom at ${\bm r}$.}
\end{figure}

The Hamiltonian of the junction system is given as
\begin{align}
    H=H_{\rm NM}+H_{\rm ph}+H_{\rm e-ph}.
\end{align}
The first term $H_{\rm NM}=\sum_{\vb*{k}\sigma}\epsilon_{\vb*{k}}c^\dagger_{\vb*{k}\sigma}c_{\vb*{k}\sigma}$ describes the electronic state in the NM, where $c^\dagger_{\vb*{k}\sigma}(c_{\vb*{k}\sigma})$ is the creation(annihilation) operator of electrons with wavenumber $\vb*{k}$, spin $\sigma$, and energy $\epsilon_{\vb*{k}}$.
The second term $H_{\rm ph} =\sum_{\vb*{q}\lambda} \hbar\omega_{\vb*{q}\lambda} (a^\dagger_{\vb*{q}\lambda}a_{\vb*{q}\lambda}+1/2)$ describes chiral phonons in the CI, where $\omega_{\vb*{q}\lambda}, a^\dagger_{\vb*{q}\lambda}(a_{\vb*{q}\lambda})$ are the frequency and the creation(annihilation) operator of phonons with wavenumber $\vb*{q}$ and circularity $\lambda$, respectively.
As schematically shown in Fig.~\ref{fig:phonon}(a), the phonon frequency varies with the circularity ($\omega_{{\bm q}+}\ne\omega_{{\bm q}-}$) because of the chirality of the crystal; there is no inversion symmetry ($\omega_{{\bm q}\lambda}\ne\omega_{-{\bm q}\lambda}$), but the time-reversal symmetry ($\omega_{{\bm q}\lambda}=\omega_{-{\bm q}\bar{\lambda}}$) exists, where $\bar{\lambda}=\mp$ denotes the circularity opposite to $\lambda$.
This feature of chiral phonons is crucial to the spin current arising at the CI-NM interface, as will be shown later.

The third term $H_{\rm e-ph}$ describes the interaction between phonons in the CI and electrons in the NM through the interface.
In this study, we focus on the effect of microscopic rotation induced by phonons as shown in Fig.~\ref{fig:phonon}(b).
This rotational motion can induce direct spin-phonon coupling through the gyromagnetic effect (the Barnett effect)~\cite{Barnett1915,Einstein-deHaas1915,Scott1962}.
This type of spin-microrotation coupling has been studied in surface acoustic waves~\cite{Kobayashi2017,Kurimune2020,tateno2021Phys.Rev.B}, laminar flows~\cite{Takahashi2020,Kazerooni2021,Kazerooni2020}
and nano-mechanical torque detection~\cite{Zolfagharkhani-NatNano-2008}, while it has been overlooked for a long time in the context of the spin-phonon coupling.
Recently, it has been shown that the electron spin in the NM couples with a microscopic phonon-mediated lattice rotation via the gyromagnetic effect in the CI~\cite{Funato2024}.
The angular velocity of the microscopic rotation is characterized by a vorticity induced by phonons, which is defined as $\vb*{\Omega}(\vb*{r})=\nabla\times\dot{\vb*{u}}(\vb*{r})$ where $\vb*{u}(\vb*{r})$ is a lattice displacement (for a schematic illustration of the vorticity $\vb*{\Omega}(\vb*{r})$, see Fig.~\ref{fig:phonon}(b)).
From the quantized form of the lattice displacement, the Fourier component of the vorticity is written as
\begin{align}
    \vb*{\Omega}_{\vb*{q}\lambda}
    =\sqrt{\frac{\hbar\omega_{\vb*{q}\lambda}}{2\rho V_{\rm CI}}}
    (\vb*{q}\times\vb*{e}_{\vb*{q}\lambda})
    \pqty{a_{\vb*{q}\lambda}
    -a^\dagger_{\vb*{q}\lambda}},
\end{align}
where $\rho$ is the mass density of the lattice,
$V_{\rm CI}$ is the volume of the CI,
$\vb*{e}_{\vb*{q}\lambda}$ are the polarization vectors,
and $\vb*{e}_{-\vb*{q}\bar{\lambda}}
    =\vb*{e}^*_{\vb*{q}\lambda}$
holds.
The effective Hamiltonian up to second order in the electron tunneling process and first order in the spin-microrotation coupling in the CI can be derived as~\cite{Funato2024}
\begin{align}
    H_{\rm e-ph}
     & =-\sum_{\vb*{p}\vb*{q}\lambda}
    J_{\vb*{q},\vb*{p}}
    \pqty{\Omega^+_{\vb*{q}\lambda}\hat{s}^-_{-\vb*{p}}
    +\Omega^-_{-\vb*{q}\bar{\lambda}}\hat{s}^+_{\vb*{p}}},
\end{align}
where $\hat{s}^\pm_{\vb*{p}} =\hat{s}^x_{\vb*{p}} \pm i\hat{s}^y_{\vb*{p}}$ is the ladder operator of the electron spin in the NM, $\hat{\vb*{s}}_{\vb*{p}}
    =(1/2)\sum_{\vb*{k}\sigma\sigma'}
    c^\dagger_{\vb*{k}-\vb*{p}\sigma}
    \vb*{\sigma}_{\sigma\sigma'}
    c_{\vb*{k}\sigma'}$,
$\vb*{\sigma}=(\sigma_x,\sigma_y,\sigma_z)$ are the Pauli matrices, and $\Omega^\pm_{\vb*{q}\lambda} =\Omega^x_{\vb*{q}\lambda} \pm i\Omega^y_{\vb*{q}\lambda}$ is the ladder operator for the phonon vorticity.

The interfacial spin current operator is defined as
\begin{align}
    \hat{I}_{\rm s}
    &\equiv-\hbar\pdv{\hat{s}^z_{\vb{0}}}{t}
    =i[\hat{s}^z_{\vb{0}},H]
    =i\sum_{\vb*{p}\vb*{q}\lambda}
    J_{\vb*{q},\vb*{p}}
    \pqty{\Omega^+_{\vb*{q}\lambda}\hat{s}^-_{-\vb*{p}}
    -\Omega^-_{-\vb*{q}\bar{\lambda}}\hat{s}^+_{\vb*{p}}}.
\end{align}
The statistical average of the spin current, $\ev*{\hat{I}_{\rm s}}$, can be calculated within a first-order perturbation of $H_{\rm e-ph}$ as
\begin{align}
    \ev*{\hat{I}_{\rm s}}
    &=\frac{4\hbar^2\abs{J}^2}
    {\rho V_{\rm CI}}
    \sum_{\vb*{p}\vb*{q}\lambda}
    \omega_{\vb*{q}\lambda}
    q_z
    [\vb*{q}\cdot\Im(\vb*{e}^*_{\vb*{q}\lambda}\times \vb*{e}_{\vb*{q}\lambda})]
    \nonumber\\
    &\times
    \int_{-\infty}^\infty
    \frac{\dd{\omega}}{2\pi}
    \Im\chi^{\rm R}_{\vb*{p}}(\omega)
    [-\Im\mathcal{D}^{\rm R}_{\vb*{q}\lambda}
    (\omega)]
    \bqty{f^{\rm ph}_{\vb*{q}\lambda}(\omega)
    -f^{\rm m}_{\vb*{k}}(\omega)},
    \label{eq:Isav}
\end{align}
where $J_{\vb*{q},\vb*{p}}$ is assumed to be a constant $J$ for simplicity, assuming a rough interface.
A detailed calculation is given in the Appendix.
Here, $\mathcal{D}^{\rm R}_{\vb*{q}\lambda}(\omega)$ and $\chi^{\rm R}_{\vb*{p}}(\omega)$ are the Fourier transformations of the retarded components for the phonon Green function of the CI and the spin susceptibilities of the NM, which are respectively defined as
\begin{align}
    \mathcal{D}^{\rm R}_{\vb*{q}\lambda}(t)
    &=-(i/\hbar)\theta(t) \ev*{[a_{\vb*{q}\lambda}(t), a^\dagger_{\vb*{q}\lambda}(0)]}_0, \\
    \chi^{\rm R}_{\vb*{p}}(t)
    &=(i/\hbar)\theta(t) \ev*{[\hat{s}^+_{\vb*{p}}(t),\hat{s}^-_{-\vb*{p}}(0)]}_0,
\end{align}
where $\theta(t)$ is a step function and $\langle \cdots \rangle_0$ denotes the statistical average for the unperturbed Hamiltonian $H_{\rm NM}+H_{\rm ph}$.
Note that $-(1/\pi)\Im\mathcal{D}^{\rm R}_{\vb*{q}\lambda}(\omega)$ and $(1/\pi)\Im\chi^{\rm R}_{\vb*{p}}(\omega)$ correspond to the spectrum functions for spin and phonon excitations and are expressed as 
\begin{align}
    -\frac{1}{\pi} \Im\mathcal{D}^{\rm R}_{\vb*{q}\lambda}(\omega) &=\delta(\hbar\omega-\hbar\omega_{\vb*{q}\lambda}) , \\
    \frac{1}{\pi} \sum_{\vb*{p}}\Im\chi^{\rm R}_{\vb*{p}}(\omega) &=\nu_{\rm F}^2N_{\rm N}^2\hbar\omega,
\end{align}
where $\nu_{\rm F}$ is the density of states at the Fermi level per unit cell in the NM and $N_{\rm N}$ is the number of the unit cells in the NM.
Equation~\eqref{eq:Isav} also includes nonequilibrium distribution functions for spin excitations in the NM and phonon excitations in the CI, i.e., $f^{\rm m}_{\vb*{k}}(\omega)$ and $f^{\rm ph}_{\vb{q}\lambda}(\omega)$, respectively.
Here, we have assumed that the conduction electrons in the NM are in thermal equilibrium at temperature $T$, indicating that the spin-excitation distribution can be described by the Bose-Einstein distribution function as $f^{\rm m}_{\vb*{k}}(\omega)=f_0(\omega,T)=(e^{\hbar\omega/k_{\rm B}T}-1)^{-1}$.

\section{Results}

The nonequilibrium distribution functions of phonons, $f^{\rm ph}_{\vb*{q}\lambda}$, can be determined by solving the Boltzmann equation within the relaxation time approximation up to the first order of the temperature gradient.
By substituting the solution of $f^{\rm ph}_{\vb*{q}\lambda}$ into the formula, Eq.~(\ref{eq:Isav}), we obtain 
\begin{align}
    \ev*{\hat{I}_{\rm s}}
    &=\frac{2\pi\hbar^3\nu_{\rm F}^2N_{\rm N}^2\abs{J}^2}
    {\rho V_{\rm CI}k_{\rm B}T}
    \pqty{-\frac{\partial_z T}{T}}
    \nonumber\\
    &\times
    \sum_{\vb*{q}\lambda}
    \tau_{\vb*{q}\lambda}
    \omega_{\vb*{q}\lambda}^3
    v^z_{\vb*{q}\lambda}
    q_z
    [\vb*{q}\cdot\Im(\vb*{e}^*_{\vb*{q}\lambda}\times \vb*{e}_{\vb*{q}\lambda})]
    \frac{e^{\hbar\omega_{\vb*{q}\lambda}/k_{\rm B}T}}
    {(e^{\hbar\omega_{\vb*{q}\lambda}/k_{\rm B}T}-1)^2},
    \label{eq:Is}
\end{align}
where $\vb*{v}_{\vb*{q}\lambda}=\partial_{\vb*{q}}\omega_{\vb*{q}\lambda}$ and $\tau_{\vb*{q}\lambda}$ are the velocity and momentum relaxation time of phonons, respectively.
Here, we stress that the temperature gradient $(-\partial_z T)$ is crucial to the generation of the spin current across the interface.

The temperature gradient can be determined by the continuous condition for heat current as follows.
The heat current in the bulk CI, $I_{\rm h}$, is expressed as
\begin{align}
    I_{\rm h}
     & =\frac{\kappa S}{L} (\Delta T-\Delta T'),
\end{align}
where $S$ is the cross-sectional area of the interface and $\kappa$ is the thermal conductivity of the CI.
The heat current through the interface $I'_{\rm h}$ can be expressed using the temperature difference at the interface $\Delta T'$ and interfacial thermal conductance $G_{\rm h}$, as follows:
\begin{align}
    I'_{\rm h}
     & =G_{\rm h}\Delta T'.
\end{align}
From the condition $I_{\rm h}=I'_{\rm h}$,
the temperature difference at the interface $\Delta T'$ can be determined as
\begin{align}
    \Delta T'
    =\frac{\kappa S/L}{G_{\rm h}+\kappa S/L}
    \Delta T.
\end{align}
Accordingly, the spin current can be rewritten as
\begin{align}
    \ev*{\hat I_{\rm s}}
    &=\frac{\pi\hbar^3\nu_{\rm F}^2N_{\rm N}^2\abs{J}^2}
    {\rho k_{\rm B}T}
    \frac{G_{\rm h}}{LG_{\rm h}+\kappa S}
    \frac{\Delta T}{T} g(T), \\
g(T) &= \frac{1}{V_{\rm CI}} 
    \sum_{\vb*{q}\in q_z>0} \sum_{\lambda} 
    d_\lambda \tau_{\vb*{q}\lambda} q_z^2 
    \frac{e^{\hbar\omega_{\vb*{q}\lambda}/k_{\rm B}T}}
    {(e^{\hbar\omega_{\vb*{q}\lambda}/k_{\rm B}T}-1)^2}
    \pdv{\omega_{\vb*{q}\lambda}^4}{q_z},
    \label{eq:Is_final} 
\end{align}
where $d_\pm = \pm 1$ indicates the sign depending on the circularity and time-reversal symmetry ($\omega_{\vb*{q}\lambda}=\omega_{-\vb*{q}\bar{\lambda}}$) has been used.
It is clear from Eq. (\ref{eq:Is_final})
that spin current is generated only
when the material has the structural chirality,
i.e., when it lacks the parity symmetry ($\omega_{\vb*{q}+}\ne\omega_{\vb*{q}-}$, see also Fig.~\ref{fig:phonon}(a)).
This is the main result of our study.

Let us discuss the features of the obtained formula.
The temperature-dependent factor $g(T)$ is controllable.
The geometrical factor $h(L)=G_{\rm h}/(LG_{\rm h}+\kappa S) = 1/(L + L_0)$ ($L_0 = \kappa S/G_{\rm h}$) represents the effect of the temperature profile in the bulk CI.
Note that $G_{\rm h}$ is proportional to the junction area $S$, and $h(L)$ is independent of $S$.
We find that the temperature gradient in the bulk CI is dominant ($\Delta T' \ll \Delta T$) for $L>L_0$, while the temperature difference at the interface is dominant ($\Delta T' \simeq \Delta T$) for $L<L_0$.
The length dependence changes in these two cases; the factor $h(L)$ is proportional to $1/L$ in the former case, while it is independent of $L$ in the latter case.
The thermal conductivity $\kappa$ is calculated using the Boltzmann equation within the relaxation time approximation as~\cite{Callaway1959}
\begin{align}
    \kappa
    =\frac{1}{V_{\rm CI}}
    \sum_{\vb*{q}\lambda}
    \tau_{\vb*{q}\lambda}
   (v^z_{\vb*{q}\lambda})^2
   \frac{(\hbar\omega_{\vb*{q}\lambda})^2}
   {k_{\rm B}T^2}
   \frac{e^{\hbar\omega_{\vb*{q}\lambda}/k_{\rm B}T}}
   {(e^{\hbar\omega_{\vb*{q}\lambda}/k_{\rm B}T}-1)^2}.
\end{align}
In general, both the thermal conductivity and the spin current driven by the temperature gradient have complex temperature dependences since the relaxation time $\tau_{{\bm q}\lambda}$ is given by a sum of several scattering processes, including boundary scattering, the normal process, and the Umklapp process, each with different temperature dependences~\cite{Ziman2001}.

\section{Discussion}

In this study, we have focused on spin-phonon coupling mediated by microrotation induced by chiral phonons.
It is worth noting that spin-phonon coupling can also arise from a different mechanism involving spin-orbit interaction.  
For instance, in systems containing heavy elements, a dominant spin-phonon coupling has been derived based on the Berry phase mechanism combined with conventional (non-rotational) electron-phonon coupling~\cite{Yao2025}.  
In contrast to this mechanism, our microrotation-based spin coupling is expected to be effective even in materials composed solely of light elements, as it does not rely on the strength of spin-orbit interaction.  
Accordingly, we expect that our theory is well suited to junctions involving chiral systems, such as chiral organic compounds or a thin organic molecular layer on the material interface, as studied in Ref.~\cite{Kim2023}.  
Furthermore, our approach can also be applied to junctions incorporating inorganic chiral crystals, such as $\alpha$-quartz~\cite{Ohe2024}.

It is worth noting that our theory also predicts spin current generation even in achiral materials.  
As confirmed by Eq.~(\ref{finalformula}) in Appendix~\ref{appA}, the spin current arises when one of the two chiral phonon modes is more strongly excited than the other ($f^{\rm ph}_{{\bm q}+}(\omega) \ne f^{\rm ph}_{{\bm q}-}(\omega)$).  
This implies that the phonon-activated spin Seebeck effect can be induced by an imbalance in the excitations of chiral phonon modes.

\section{Summary}

We theoretically formulated 
a phonon-mediated spin current induced by a temperature gradient in a junction composed of a chiral insulator (CI) and a normal metal (NM). Our analysis revealed that gyromagnetic coupling in the CI, arising from microscopic lattice rotation induced by chiral phonons, plays a key role in generating the spin current. These results highlight the critical roles of both the temperature gradient and the non-equilibrium phonon distribution in the CI in driving spin current generation. 
Our formula also indicates that there is a characteristic length, at which the length dependence of the spin current changes.
Importantly, our proposed mechanism enables phonon-mediated spin-current generation without relying on spin-orbit interactions, offering a pathway toward chirality-driven spintronic devices composed solely of light elements.

In this work, we focused on acoustic phonons, which are expected to contribute dominantly to spin current generation at low temperatures. It is worth noting that optical phonons may have a more significant impact when experiments are conducted at room temperature, as reported in Ref.~\cite{Kim2023}. Extending the spin-microrotation coupling framework to optical phonons remains an important direction for future research. Another crucial challenge is to quantitatively estimate the spin current amplitude and compare it with theoretical predictions based on spin-orbit interactions~\cite{Ohe2024,Yao2024,Yao2025}. Furthermore, elucidating the relationship between our findings and chirality-induced spin selectivity (CISS), which was originally observed in DNA and peptides~\cite{Goehler2011,Naaman2012,Michaeli2016,Naaman2019,Mishra2020,Sano2024} and has recently been applied to spintronic devices~\cite{Miwa2020,Kondou2022,Kondou2023,Miwa2024}, remains a compelling and open problem.

\section*{ACKNOWLEDGMENTS}

This work was supported by the National Natural Science Foundation of China (NSFC) under Grant No. 12374126, 
by the Priority Program of the Chinese Academy of Sciences under Grant No. XDB28000000, and by JSPS KAKENHI under Grants (Nos. JP21H01800, JP21H04565, JP23H01839, JP24K06951, JP24H00322 and JP25KJ0400) from MEXT, Japan.

\appendix

\section{Detailed Derivation of Spin Current}
\label{appA}

In this Appendix, we provide a detailed derivation of the spin current at an interface, i.e., Eq.~\eqref{eq:Isav} in the main text, on the basis of the method of the tunneling Hamiltonian~\cite{Funato2024,Bruus2004}.
We start with an expression for the average spin current:
\begin{align}
    \ev*{\hat{I}_{\rm s}(t)}
    =\Re\bqty{2i\sum_{\vb*{p}\vb*{q}\lambda}
    J_{\vb*{q},\vb*{p}}
    \ev*{T_K\Omega^+_{\vb*{q}\lambda}(\tau_1)
    \hat{s}^-_{-\vb*{p}}(\tau_2)}},
\end{align}
where $T_K$ is the time-ordering operator on the Keldysh contour
and $\tau_1=(t,+)$ and $ \tau_2=(t,-)$ are the time variables on the forward and backward paths,
respectively.
The statistical average can be calculated as a perturbative expansion in $H_{\rm e-ph}$:
\begin{align}
    \ev*{\hat{I}_{\rm s}(t)}
    =\Re\bqty{2i\sum_{\vb*{p}\vb*{q}\lambda}
    J_{\vb*{q},\vb*{p}}
    \ev{T_K\hat{s}^-_{-\vb*{p}}(\tau_2)
    \Omega^+_{\vb*{q}\lambda}(\tau_1)
    \exp\pqty{-\frac{i}{\hbar}
        \int_C\dd{\tau}\hat{\mathcal{H}}_{\rm e-ph}(\tau)}}_0},
\end{align}
where $C$ is the Keldysh contour and the integral is taken over the Keldysh time $\tau=(t',\eta)$ with $\eta=\pm$.
$\langle \cdots \rangle_0$ denotes a statistical average for the unperturbed Hamiltonian $H_{\rm NM}+H_{\rm ph}$.
Within a first-order perturbation of $H_{\rm e-ph}$ we obtain:
\begin{align}
    \ev*{\hat{I}_{\rm s}(t)}
     & =-
    \frac{2}{\hbar}\int_C\dd{\tau}
    \Re\bqty{\sum_{\vb*{p}\vb*{q}\lambda}
    \abs{J_{\vb*{q},\vb*{p}}}^2
    \ev*{T_K\hat{s}^+_{\vb*{p}}(\tau)
    \hat{s}^-_{-\vb*{p}}(\tau_2)}_0
    \ev*{T_K
    \Omega^+_{\vb*{q}\lambda}(\tau_1)
    \Omega^-_{-\vb*{q}\bar{\lambda}}(\tau)}_0}.
\end{align}
By substituting into this equation the definition of vorticity $\vb*{\Omega}_{\vb*{q}\lambda}$ given in Eq.~(2) in the main text,
we obtain
\begin{align}
    \ev*{\hat{I}_{\rm s}(t)}
    &=\frac{\hbar^2}
    {\rho V_{\rm CI}}
    \int_C\dd{\tau}
    \sum_{\vb*{p}\vb*{q}\lambda}
    \abs{J_{\vb*{q},\vb*{p}}}^2
    \omega_{\vb*{q}\lambda}
    (\vb*{q}\times\vb*{e}_{\vb*{q}\lambda})_+
    (\vb*{q}\times\vb*{e}^*_{\vb*{q}\lambda})_-
    \nonumber\\
    &\hspace{5mm} \times\Re\Bqty{
    \chi_{\vb*{p}}
    (\tau-\tau_2)
    \bqty{\mathcal{D}_{\vb*{q}\lambda}
    (\tau_1-\tau)
    +\mathcal{D}_{-\vb*{q}\bar{\lambda}}
    (\tau-\tau_1)}},
\end{align}
where 
\begin{align}
\mathcal{D}_{\vb*{q}\lambda}
    (\tau_1-\tau_2)
     & =-(i/\hbar)\ev*{T_K
        a_{\vb*{q}\lambda}(\tau_1)
    a^\dagger_{\vb*{q}\lambda}(\tau_2)}_0 , \\
\chi_{\vb*{p}}
    (\tau_1-\tau_2)
    & =(i/\hbar)\ev*{T_K
    \hat{s}^+_{\vb*{p}}(\tau_1)
    \hat{s}^-_{-\vb*{p}}(\tau_2)}_0,
\end{align}
are the phonon Green function of the CI and the spin susceptibilities of the NM,
respectively.
Since the integral on the Keldysh contour $C$ is defined as $\int_C\dd{\tau}\cdots=\int_{-\infty}^\infty\dd{t'}\sum_{\eta=\pm}\eta\cdots$,
we expand the summation over $\eta$ and assume that $J_{\vb*{q},\vb*{p}}$ is a constant $J$ for simplicity.
Accordingly, we find that
\begin{align}
    \ev*{\hat{I}_{\rm s}(t)}
    &=\frac{\hbar^2\abs{J}^2}
    {\rho V_{\rm CI}}
    \int_{-\infty}^\infty\dd{t'}
    \sum_{\vb*{p}\vb*{q}\lambda}
    \omega_{\vb*{q}\lambda}
    (\vb*{q}\times\vb*{e}_{\vb*{q}\lambda})_+
    (\vb*{q}\times\vb*{e}^*_{\vb*{q}\lambda})_- \nonumber\\
    &\hspace{5mm} \times\Re\Big[
    \mathcal{D}^{\rm R}_{\vb*{q}\lambda}
    (t-t')
    \chi^{<}_{\vb*{p}}
    (t'-t)
    +\mathcal{D}^{<}_{\vb*{q}\lambda}
    (t-t')
    \chi^{\rm A}_{\vb*{p}}
    (t'-t)                                    \nonumber\\
    &\hspace{10mm}+\mathcal{D}^{\rm A}_{-\vb*{q}\bar{\lambda}}
    (t'-t)
    \chi^{<}_{\vb*{p}}
    (t'-t)
    +\mathcal{D}^{>}_{-\vb*{q}\bar{\lambda}}
    (t'-t)
    \chi^{\rm A}_{\vb*{p}}
    (t'-t)\Big]\nonumber\\
    &=\frac{\hbar^2\abs{J}^2}
    {\rho V_{\rm CI}}
    \int_{-\infty}^\infty
    \frac{\dd{\omega}}{2\pi}
    \sum_{\vb*{p}\vb*{q}\lambda}
    \omega_{\vb*{q}\lambda}
    (\vb*{q}\times\vb*{e}_{\vb*{q}\lambda})_+
    (\vb*{q}\times\vb*{e}^*_{\vb*{q}\lambda})_- \nonumber\\
    &\hspace{5mm}\times\Re\Big[
    \mathcal{D}^{\rm R}_{\vb*{q}\lambda}
    (\omega)
    \chi^{<}_{\vb*{p}}
    (\omega)
    +\mathcal{D}^{<}_{\vb*{q}\lambda}
    (\omega)
    \chi^{\rm A}_{\vb*{p}}
    (\omega)                                    \nonumber\\
    &\hspace{10mm}+\mathcal{D}^{\rm A}_{-\vb*{q}\bar{\lambda}}
    (-\omega)
    \chi^{<}_{\vb*{p}}
    (\omega)
    +\mathcal{D}^{>}_{-\vb*{q}\bar{\lambda}}
    (-\omega)
    \chi^{\rm A}_{\vb*{p}}
    (\omega)\Big]\nonumber\\
    &=-\frac{\hbar^2\abs{J}^2}
    {\rho V_{\rm CI}}
    \int_{-\infty}^\infty
    \frac{\dd{\omega}}{2\pi}
    \sum_{\vb*{p}\vb*{q}\lambda}
    \omega_{\vb*{q}\lambda}
    (\vb*{q}\times\vb*{e}_{\vb*{q}\lambda})_+
    (\vb*{q}\times\vb*{e}^*_{\vb*{q}\lambda})_- \nonumber\\
    &\hspace{5mm}\times\Big[
    \Im\pqty{\mathcal{D}^{\rm R}_{\vb*{q}\lambda}
    (\omega)
    +\mathcal{D}^{\rm A}_{-\vb*{q}\bar{\lambda}}
    (-\omega)}
    \Im\chi^{<}_{\vb*{p}}
    (\omega)\nonumber\\
    &\hspace{10mm}+\Im\pqty{\mathcal{D}^{<}_{\vb*{q}\lambda}
    (\omega)
    +\mathcal{D}^{>}_{-\vb*{q}\bar{\lambda}}
    (-\omega)}
    \Im\chi^{\rm A}_{\vb*{p}}
    (\omega)\Big].
\end{align}
We define the nonequilibrium distribution functions for spin excitation in the NM and phonons in the CI as
\begin{align}
f^{\rm m}_{\vb{p}}(\omega)
    &=\chi^<_{\vb{p}}(\omega)
    /2i\Im\chi^{\rm R}_{\vb{p}}(\omega), \\
f^{\rm ph}_{\vb{q}\lambda}(\omega) &=\mathcal{D}^<_{\vb{q}\lambda}(\omega)
    /2i\Im\mathcal{D}^{\rm R}_{\vb{q}\lambda}(\omega).
\end{align}
Using these definitions, we obtain
\begin{align}
    \ev*{\hat{I}_{\rm s}(t)}
    &=-\frac{2\hbar^2\abs{J}^2}
    {\rho V_{\rm CI}}
    \int_{-\infty}^\infty
    \frac{\dd{\omega}}{2\pi}
    \sum_{\vb*{p}\vb*{q}\lambda}
    \omega_{\vb*{q}\lambda}
    (\vb*{q}\times\vb*{e}_{\vb*{q}\lambda})_+
    (\vb*{q}\times\vb*{e}^*_{\vb*{q}\lambda})_- \nonumber\\
    &\hspace{5mm}\times\Im\chi^{\rm R}_{\vb*{p}}
    (\omega)
    \Big[\Im\mathcal{D}^{\rm R}_{\vb*{q}\lambda}
    (\omega)
    \Bqty{f^{\rm m}_{\vb*{p}}(\omega)
    -f^{\rm ph}_{\vb*{q}\lambda}(\omega)}\nonumber\\
    &\hspace{10mm}-\Im\mathcal{D}^{\rm R}_{-\vb*{q}\bar{\lambda}}
    (-\omega)
    \Bqty{f^{\rm m}_{\vb*{p}}(\omega)
    +\pqty{1+f^{\rm ph}_{-\vb*{q}\bar{\lambda}}(-\omega)}}
    \Big]\nonumber\\
    &=-\frac{2\hbar^2\abs{J}^2}
    {\rho V_{\rm CI}}
    \int_{-\infty}^\infty
    \frac{\dd{\omega}}{2\pi}
    \sum_{\vb*{p}\vb*{q}\lambda}
    \omega_{\vb*{q}\lambda}
    (\vb*{q}\times\vb*{e}_{\vb*{q}\lambda})_+
    (\vb*{q}\times\vb*{e}^*_{\vb*{q}\lambda})_- \nonumber\\
    &\hspace{5mm}\times
    \Big[\Im\chi^{\rm R}_{\vb*{p}}
    (\omega)\Im\mathcal{D}^{\rm R}_{\vb*{q}\lambda}
    (\omega)
    \Bqty{f^{\rm m}_{\vb*{p}}(\omega)
    -f^{\rm ph}_{\vb*{q}\lambda}(\omega)}\nonumber\\
    &\hspace{10mm}-\Im\chi^{\rm R}_{\vb*{p}}
    (-\omega)\Im\mathcal{D}^{\rm R}_{-\vb*{q}\bar{\lambda}}
    (\omega)
    \Bqty{f^{\rm m}_{\vb*{p}}(-\omega)
    +\pqty{1+f^{\rm ph}_{-\vb*{q}\bar{\lambda}}(\omega)}}
    \Big].
\end{align}
For simplicity,
we approximate $f^m_{\vb{p}}(\omega)=f_0(\omega, T)=(e^{\hbar\omega/k_{\rm B}T}-1)^{-1}$.
According, we find that
\begin{align}
    \ev*{\hat{I}_{\rm s}(t)}
    &=-\frac{2\hbar^2\abs{J}^2}
    {\rho V_{\rm CI}}
    \int_{-\infty}^\infty
    \frac{\dd{\omega}}{2\pi}
    \sum_{\vb*{p}\vb*{q}\lambda}
    \omega_{\vb*{q}\lambda}
    (\vb*{q}\times\vb*{e}_{\vb*{q}\lambda})_+
    (\vb*{q}\times\vb*{e}^*_{\vb*{q}\lambda})_- \nonumber\\
    &\hspace{5mm}\times
    \Im\chi^{\rm R}_{\vb*{p}}
    (\omega)
    \Big\{[-\Im\mathcal{D}^{\rm R}_{\vb*{q}\lambda}
    (\omega)]
    \bqty{f^{\rm ph}_{\vb*{q}\lambda}(\omega)-f_0(\omega, T)}\nonumber\\
    &\hspace{10mm}-[-\Im\mathcal{D}^{\rm R}_{-\vb*{q}\bar{\lambda}}
    (\omega)]
    \bqty{f^{\rm ph}_{-\vb*{q}\bar{\lambda}}(\omega)-f_0(\omega, T)}
    \Big\}.
\end{align}
Here,
we have used $\Im\chi^{\rm R}_{\vb*{p}}(-\omega)=-\Im\chi^{\rm R}_{\vb*{p}}(\omega)$.
Finally, we obtain
\begin{align}
    \ev*{\hat{I}_{\rm s}}
    &=\frac{4\hbar^2\abs{J}^2}
    {\rho V_{\rm CI}}
    \sum_{\vb*{p}\vb*{q}\lambda}
    \omega_{\vb*{q}\lambda}
    q_z
    [\vb*{q}\cdot\Im(\vb*{e}^*_{\vb*{q}\lambda}*\vb*{e}_{\vb*{q}\lambda})]
    \nonumber\\
    &\hspace{5mm}\times
    \int_{-\infty}^\infty
    \frac{\dd{\omega}}{2\pi}
    \Im\chi^{\rm R}_{\vb*{p}}(\omega)
    [-\Im\mathcal{D}^{\rm R}_{\vb*{q}\lambda}
    (\omega)]
    [f^{\rm ph}_{\vb*{q}\lambda}(\omega)
    -f_0(\omega, T)]. \label{finalformula}
\end{align}

\bibliographystyle{elsarticle-num} 
\bibliography{references}

\begin{thebibliography}{10}
\expandafter\ifx\csname url\endcsname\relax
  \def\url#1{\texttt{#1}}\fi
\expandafter\ifx\csname urlprefix\endcsname\relax\def\urlprefix{URL }\fi
\expandafter\ifx\csname href\endcsname\relax
  \def\href#1#2{#2} \def\path#1{#1}\fi

\bibitem{Bauer2012}
G.~E.~W. Bauer, E.~Saitoh, B.~J. van Wees, Spin caloritronics, Nat. Mater. 11 (2012) 391--399.
\newblock \href {https://doi.org/10.1038/nmat3301} {\path{doi:10.1038/nmat3301}}.

\bibitem{Hoffmann2015}
A.~Hoffmann, S.~D. Bader, {Opportunities at the Frontiers of Spintronics}, Phys. Rev. Appl. 4 (2015) 047001.
\newblock \href {https://doi.org/10.1103/PhysRevApplied.4.047001} {\path{doi:10.1103/PhysRevApplied.4.047001}}.

\bibitem{Uchida2021}
K.-i. Uchida, R.~Iguchi, {Spintronic Thermal Management}, J. Phys. Soc. Jpn. 90 (2021) 122001.
\newblock \href {https://doi.org/10.7566/JPSJ.90.122001} {\path{doi:10.7566/JPSJ.90.122001}}.

\bibitem{Uchida2008}
K.~Uchida, S.~Takahashi, K.~Harii, J.~Ieda, W.~Koshibae, K.~Ando, S.~Maekawa, E.~Saitoh, {Observation of the spin Seebeck effect}, Nature 455 (2008) 778--781.
\newblock \href {https://doi.org/10.1038/nature07321} {\path{doi:10.1038/nature07321}}.

\bibitem{Jaworski2010}
C.~M. Jaworski, J.~Yang, S.~Mack, D.~D. Awschalom, J.~P. Heremans, R.~C. Myers, {Observation of the spin-Seebeck effect in a ferromagnetic semiconductor}, Nat. Mater. 9 (2010) 898--903.
\newblock \href {https://doi.org/10.1038/nmat2860} {\path{doi:10.1038/nmat2860}}.

\bibitem{Slachter2010}
A.~Slachter, F.~L. Bakker, J.-P. Adam, B.~J. van Wees, {Thermally driven spin injection from a ferromagnet into a non-magnetic metal}, Nat. Phys. 6 (2010) 879--882.
\newblock \href {https://doi.org/10.1038/nphys1767} {\path{doi:10.1038/nphys1767}}.

\bibitem{Uchida2011}
K.~Uchida, H.~Adachi, T.~An, T.~Ota, M.~Toda, B.~Hillebrands, S.~Maekawa, E.~Saitoh, {Long-range spin Seebeck effect and acoustic spin pumping}, Nat. Mater. 10 (2011) 737--741.
\newblock \href {https://doi.org/10.1038/nmat3099} {\path{doi:10.1038/nmat3099}}.

\bibitem{Jaworski2012}
C.~M. Jaworski, R.~C. Myers, E.~Johnston-Halperin, J.~P. Heremans, {Giant spin Seebeck effect in a non-magnetic material}, Nature 487 (2012) 210--213.
\newblock \href {https://doi.org/10.1038/nature11221} {\path{doi:10.1038/nature11221}}.

\bibitem{Adachi2013}
H.~Adachi, K.-I. Uchida, E.~Saitoh, S.~Maekawa, {Theory of the spin Seebeck effect}, Rep. Prog. Phys. 76 (2013) 036501.
\newblock \href {https://doi.org/10.1088/0034-4885/76/3/036501} {\path{doi:10.1088/0034-4885/76/3/036501}}.

\bibitem{Kim2023}
K.~Kim, E.~Vetter, L.~Yan, C.~Yang, Z.~Wang, R.~Sun, Y.~Yang, A.~H. Comstock, X.~Li, J.~Zhou, L.~Zhang, W.~You, D.~Sun, J.~Liu, {Chiral-phonon-activated spin Seebeck effect}, Nat. Mater. 22 (2023) 322--328.
\newblock \href {https://doi.org/10.1038/s41563-023-01473-9} {\path{doi:10.1038/s41563-023-01473-9}}.

\bibitem{Barron2012}
L.~D. Barron, {From Cosmic Chirality to Protein Structure: Lord Kelvin's Legacy}, Chirality 24 (2012) 879--893.
\newblock \href {https://doi.org/https://doi.org/10.1002/chir.22017} {\path{doi:https://doi.org/10.1002/chir.22017}}.

\bibitem{Kishine2022}
J.-i. Kishine, H.~Kusunose, H.~M. Yamamoto, {On the Definition of Chirality and Enantioselective Fields}, Isr. J. Chem. 62 (2022) e202200049.
\newblock \href {https://doi.org/https://doi.org/10.1002/ijch.202200049} {\path{doi:https://doi.org/10.1002/ijch.202200049}}.

\bibitem{Fransson2022}
J.~Fransson, {The Chiral Induced Spin Selectivity Effect What It Is, What It Is Not, And Why It Matters}, Isr. J. Chem. 62 (2022) e202200046.
\newblock \href {https://doi.org/https://doi.org/10.1002/ijch.202200046} {\path{doi:https://doi.org/10.1002/ijch.202200046}}.

\bibitem{Zhang2025b}
S.~Zhang, Z.~Huang, M.~Du, T.~Ying, L.~Du, T.~Zhang, {The chirality of phonons: Definitions, symmetry constraints, and experimental observation}, arXiv:2503.22794 (2025).
\newblock \href {https://doi.org/10.48550/arXiv.2503.22794} {\path{doi:10.48550/arXiv.2503.22794}}.

\bibitem{Zhang2025}
T.~T. Zhang, Y.~Liu, H.~Miao, S.~Murakami, {New Advances in Phonons: From Band Topology to Quasiparticle Chirality}, arXiv:2505.06179 (2025).
\newblock \href {https://doi.org/10.48550/arXiv.2505.06179} {\path{doi:10.48550/arXiv.2505.06179}}.

\bibitem{Zhang-Niu2014}
L.~Zhang, Q.~Niu, {Angular Momentum of Phonons and the Einstein--de Haas Effect}, Phys. Rev. Lett. 112 (2014) 085503.
\newblock \href {https://doi.org/10.1103/PhysRevLett.112.085503} {\path{doi:10.1103/PhysRevLett.112.085503}}.

\bibitem{Vonsovskii1962}
S.~V. Vonsovskii, M.~S. Svirskii, {Phonon Spin}, Sov. Phys. Solid State 3 (1962) 1568.

\bibitem{Garanin-Chudnovsky2015}
D.~A. Garanin, E.~M. Chudnovsky, {Angular momentum in spin-phonon processes}, Phys. Rev. B 92 (2015) 024421.
\newblock \href {https://doi.org/10.1103/PhysRevB.92.024421} {\path{doi:10.1103/PhysRevB.92.024421}}.

\bibitem{Nakane-Kohno2018}
J.~J. Nakane, H.~Kohno, Angular momentum of phonons and its application to single-spin relaxation, Phys. Rev. B 97 (2018) 174403.
\newblock \href {https://doi.org/10.1103/PhysRevB.97.174403} {\path{doi:10.1103/PhysRevB.97.174403}}.

\bibitem{Geilhufe2022}
R.~M. Geilhufe, Dynamic electron-phonon and spin-phonon interactions due to inertia, Phys. Rev. Res. 4 (2022) L012004.
\newblock \href {https://doi.org/10.1103/PhysRevResearch.4.L012004} {\path{doi:10.1103/PhysRevResearch.4.L012004}}.

\bibitem{Geilhufe2023}
R.~M. Geilhufe, W.~Hergert, {Electron magnetic moment of transient chiral phonons in ${\mathrm{KTaO}}_{3}$}, Phys. Rev. B 107 (2023) L020406.
\newblock \href {https://doi.org/10.1103/PhysRevB.107.L020406} {\path{doi:10.1103/PhysRevB.107.L020406}}.

\bibitem{Basini2024}
M.~Basini, M.~Pancaldi, B.~Wehinger, M.~Udina, V.~Unikandanunni, T.~Tadano, M.~C. Hoffmann, A.~V. Balatsky, S.~Bonetti, {Terahertz electric-field-driven dynamical multiferroicity in SrTiO${}_3$}, Nature 628 (2024) 534--539.
\newblock \href {https://doi.org/10.1038/s41586-024-07175-9} {\path{doi:10.1038/s41586-024-07175-9}}.

\bibitem{Davies2024}
C.~S. Davies, F.~G.~N. Fennema, A.~Tsukamoto, I.~Razdolski, A.~V. Kimel, A.~Kirilyuk, {Phononic switching of magnetization by the ultrafast Barnett effect}, Nature 628 (2024) 540--544.
\newblock \href {https://doi.org/10.1038/s41586-024-07200-x} {\path{doi:10.1038/s41586-024-07200-x}}.

\bibitem{Choi2024}
I.~H. Choi, S.~G. Jeong, S.~Song, S.~Park, D.~B. Shin, W.~S. Choi, J.~S. Lee, Real-time dynamics of angular momentum transfer from spin to acoustic chiral phonon in oxide heterostructures, Nat. Nanotechnol. 19 (2024) 1277--1282.
\newblock \href {https://doi.org/10.1038/s41565-024-01719-w} {\path{doi:10.1038/s41565-024-01719-w}}.

\bibitem{Juraschek2019}
D.~M. Juraschek, N.~A. Spaldin, Orbital magnetic moments of phonons, Phys. Rev. Mater. 3 (2019) 064405.
\newblock \href {https://doi.org/10.1103/PhysRevMaterials.3.064405} {\path{doi:10.1103/PhysRevMaterials.3.064405}}.

\bibitem{Xiong2022a}
G.~Xiong, H.~Chen, D.~Ma, L.~Zhang, {Effective magnetic fields induced by chiral phonons}, Phys. Rev. B 106 (2022) 144302.
\newblock \href {https://doi.org/10.1103/PhysRevB.106.144302} {\path{doi:10.1103/PhysRevB.106.144302}}.

\bibitem{Xiong2022b}
G.~Xiong, Z.~Yu, L.~Zhang, {Interband chiral phonon transfer in a magnetic field}, Phys. Rev. B 105 (2022) 024312.
\newblock \href {https://doi.org/10.1103/PhysRevB.105.024312} {\path{doi:10.1103/PhysRevB.105.024312}}.

\bibitem{Streib2018}
S.~Streib, H.~Keshtgar, G.~E.~W. Bauer, {Damping of Magnetization Dynamics by Phonon Pumping}, Phys. Rev. Lett. 121 (2018) 027202.
\newblock \href {https://doi.org/10.1103/PhysRevLett.121.027202} {\path{doi:10.1103/PhysRevLett.121.027202}}.

\bibitem{Chen2022}
H.~Chen, W.~Wu, J.~Zhu, Z.~Yang, W.~Gong, W.~Gao, S.~A. Yang, L.~Zhang, {Chiral Phonon Diode Effect in Chiral Crystals}, Nano Lett. 22 (2022) 1688--1693.
\newblock \href {https://doi.org/10.1021/acs.nanolett.1c04705} {\path{doi:10.1021/acs.nanolett.1c04705}}.

\bibitem{Suzuki2024}
Y.~Suzuki, S.~Sumita, Y.~Kato, {Theory of phonon angular momentum transport across smooth interfaces between crystals}, arXiv:2409.08874 (2024).
\newblock \href {https://doi.org/10.48550/arXiv.2409.08874} {\path{doi:10.48550/arXiv.2409.08874}}.

\bibitem{Hamada2018}
M.~Hamada, E.~Minamitani, M.~Hirayama, S.~Murakami, Phonon angular momentum induced by the temperature gradient, Phys. Rev. Lett. 121 (2018) 175301.
\newblock \href {https://doi.org/10.1103/PhysRevLett.121.175301} {\path{doi:10.1103/PhysRevLett.121.175301}}.

\bibitem{HZhang2025}
H.~Zhang, N.~Peshcherenko, F.~Yang, T.~Z. Ward, P.~Raghuvanshi, L.~Lindsay, C.~Felser, Y.~Zhang, J.~Q. Yan, H.~Miao, Measurement of phonon angular momentum, Nat. Phys. (2025).
\newblock \href {https://doi.org/10.1038/s41567-025-02952-3} {\path{doi:10.1038/s41567-025-02952-3}}.

\bibitem{Zhang-Niu2015}
L.~Zhang, Q.~Niu, {Chiral Phonons at High-Symmetry Points in Monolayer Hexagonal Lattices}, Phys. Rev. Lett. 115 (2015) 115502.
\newblock \href {https://doi.org/10.1103/PhysRevLett.115.115502} {\path{doi:10.1103/PhysRevLett.115.115502}}.

\bibitem{Chen2018}
H.~Chen, W.~Zhang, Q.~Niu, L.~Zhang, Chiral phonons in two-dimensional materials, 2D Materials 6 (2018) 012002.
\newblock \href {https://doi.org/10.1088/2053-1583/aaf292} {\path{doi:10.1088/2053-1583/aaf292}}.

\bibitem{Ren2021}
Y.~Ren, C.~Xiao, D.~Saparov, Q.~Niu, {Phonon Magnetic Moment from Electronic Topological Magnetization}, Phys. Rev. Lett. 127 (2021) 186403.
\newblock \href {https://doi.org/10.1103/PhysRevLett.127.186403} {\path{doi:10.1103/PhysRevLett.127.186403}}.

\bibitem{Saparov2022}
D.~Saparov, B.~Xiong, Y.~Ren, Q.~Niu, {Lattice dynamics with molecular Berry curvature: Chiral optical phonons}, Phys. Rev. B 105 (2022) 064303.
\newblock \href {https://doi.org/10.1103/PhysRevB.105.064303} {\path{doi:10.1103/PhysRevB.105.064303}}.

\bibitem{Yao2025}
D.~Yao, S.~Murakami, Theory of spin magnetization driven by chiral phonons, Phys. Rev. B 111 (2025) 134414.
\newblock \href {https://doi.org/10.1103/PhysRevB.111.134414} {\path{doi:10.1103/PhysRevB.111.134414}}.

\bibitem{Yao2022}
D.~Yao, S.~Murakami, Chiral-phonon-induced current in helical crystals, Phys. Rev. B 105 (2022) 184412.
\newblock \href {https://doi.org/10.1103/PhysRevB.105.184412} {\path{doi:10.1103/PhysRevB.105.184412}}.

\bibitem{Yao2024}
D.~Yao, M.~Matsuo, T.~Yokoyama, Electric field-induced nonreciprocal spin current due to chiral phonons in chiral-structure superconductors, Appl. Phys. Lett. 124 (2024).
\newblock \href {https://doi.org/10.1063/5.0207915} {\path{doi:10.1063/5.0207915}}.

\bibitem{Kato2024}
Y.~Kato, Y.~Suzuki, T.~Sato, H.~M. Yamamoto, Y.~Togawa, H.~K. Kusunose, J.-i. Koshine, Chirality-dependent spin polarization in diffusive metals: linear and quadratic responses, arXiv:2408.04450 (2024).
\newblock \href {https://doi.org/10.48550/arXiv.2408.04450} {\path{doi:10.48550/arXiv.2408.04450}}.

\bibitem{McLellan1988}
A.~G. McLellan, Angular momentum states for phonons and a rotationally invariant development of lattice dynamics, J. Phys. C: Solid State Phys. 21 (1988) 1177.
\newblock \href {https://doi.org/10.1088/0022-3719/21/7/009} {\path{doi:10.1088/0022-3719/21/7/009}}.

\bibitem{Bozovic1984}
I.~Bo\ifmmode~\check{z}\else \v{z}\fi{}ovic, Possible band-structure shapes of quasi-one-dimensional solids, Phys. Rev. B 29 (1984) 6586--6599.
\newblock \href {https://doi.org/10.1103/PhysRevB.29.6586} {\path{doi:10.1103/PhysRevB.29.6586}}.

\bibitem{Tatsumi2018}
Y.~Tatsumi, T.~Kaneko, R.~Saito, Conservation law of angular momentum in helicity-dependent raman and rayleigh scattering, Phys. Rev. B 97 (2018) 195444.
\newblock \href {https://doi.org/10.1103/PhysRevB.97.195444} {\path{doi:10.1103/PhysRevB.97.195444}}.

\bibitem{Zhang2022}
T.~Zhang, S.~Murakami, {Chiral phonons and pseudoangular momentum in nonsymmorphic systems}, Phys. Rev. Res. 4 (2022) L012024.
\newblock \href {https://doi.org/10.1103/PhysRevResearch.4.L012024} {\path{doi:10.1103/PhysRevResearch.4.L012024}}.

\bibitem{Komiyama2022}
H.~Komiyama, T.~Zhang, S.~Murakami, {Physics of phonons in systems with approximate screw symmetry}, Phys. Rev. B 106 (2022) 184104.
\newblock \href {https://doi.org/10.1103/PhysRevB.106.184104} {\path{doi:10.1103/PhysRevB.106.184104}}.

\bibitem{AKato2022}
A.~Kato, H.~M. Yamamoto, J.-i. Kishine, {Chirality-induced spin filtering in pseudo Jahn-Teller molecules}, Phys. Rev. B 105 (2022) 195117.
\newblock \href {https://doi.org/10.1103/PhysRevB.105.195117} {\path{doi:10.1103/PhysRevB.105.195117}}.

\bibitem{Tsunetsugu2023}
H.~Tsunetsugu, H.~Kusunose, {Theory of Energy Dispersion of Chiral Phonons}, J. Phys. Soc. Jpn. 92 (2023) 023601.
\newblock \href {https://doi.org/10.7566/JPSJ.92.023601} {\path{doi:10.7566/JPSJ.92.023601}}.

\bibitem{AKato2023}
A.~Kato, J.-i. Kishine, {Note on Angular Momentum of Phonons in Chiral Crystals}, J. Phys. Soc. Jpn. 92 (2023) 075002.
\newblock \href {https://doi.org/10.7566/JPSJ.92.075002} {\path{doi:10.7566/JPSJ.92.075002}}.

\bibitem{Tateishi2025}
T.~Tateishi, A.~Kato, J.-i. Kishine, {Electron–Chiral Phonon Coupling, Crystal Angular Momentum, and Phonon Chirality}, J. Phys. Soc. Jpn. 94 (2025) 053601.
\newblock \href {https://doi.org/10.7566/JPSJ.94.053601} {\path{doi:10.7566/JPSJ.94.053601}}.

\bibitem{Zhu2018}
H.~Zhu, J.~Yi, M.~Li, J.~Xiao, L.~Zhang, C.~Yang, R.~Kaindl, L.~L.J., Y.~Wang, Z.~X., Observation of chiral phonons, Science 359 (2018) 579.
\newblock \href {https://doi.org/10.1126/science.aar2711} {\path{doi:10.1126/science.aar2711}}.

\bibitem{Jeong2022}
S.~G. Jeong, J.~Kim, A.~Seo, S.~Park, H.~Y. Jeong, Y.-M. Kim, V.~Lauter, T.~Egami, J.~H. Han, W.~S. Choi, Unconventional interlayer exchange coupling via chiral phonons in synthetic magnetic oxide heterostructures, Sci. Adv. 8 (2022) eabm4005.
\newblock \href {https://doi.org/10.1126/sciadv.abm4005} {\path{doi:10.1126/sciadv.abm4005}}.

\bibitem{Ishito2023a}
K.~Ishito, H.~Mao, Y.~Kousaka, Y.~Togawa, S.~Iwasaki, T.~Zhang, S.~Murakami, J.-i. Kishine, T.~Satoh, {Truly chiral phonons in $\alpha$-HgS}, Nat. Phys. 19 (2023) 35--39.
\newblock \href {https://doi.org/10.1038/s41567-022-01790-x} {\path{doi:10.1038/s41567-022-01790-x}}.

\bibitem{Ishito2023b}
K.~Ishito, H.~Mao, K.~Kobayashi, Y.~Kousaka, Y.~Togawa, H.~Kusunose, J.-i. Kishine, T.~Satoh, {Chiral phonons: circularly polarized Raman spectroscopy and ab initio calculations in a chiral crystal tellurium}, Chirality 2023 (2023) 23544.
\newblock \href {https://doi.org/https://doi.org/10.1002/chir.23544} {\path{doi:https://doi.org/10.1002/chir.23544}}.

\bibitem{Barnett1915}
S.~J. Barnett, {Magnetization by Rotation}, Phys. Rev. 6 (1915) 239.
\newblock \href {https://doi.org/10.1103/PhysRev.6.239} {\path{doi:10.1103/PhysRev.6.239}}.

\bibitem{Einstein-deHaas1915}
A.~Einstein, W.~J. de~Haas, {Experimental proof of the existence of Amp\`{e}re's molecular currents}, KNAW proc. 18 I (1915) 696.

\bibitem{Scott1962}
G.~G. Scott, {Review of Gyromagnetic Ratio Experiments}, Rev. Mod. Phys. 34 (1962) 102.
\newblock \href {https://doi.org/10.1103/RevModPhys.34.102} {\path{doi:10.1103/RevModPhys.34.102}}.

\bibitem{Kobayashi2017}
D.~Kobayashi, T.~Yoshikawa, M.~Matsuo, R.~Iguchi, S.~Maekawa, E.~Saitoh, Y.~Nozaki, {Spin Current Generation Using a Surface Acoustic Wave Generated via Spin-Rotation Coupling}, Phys. Rev. Lett. 119 (2017) 077202.
\newblock \href {https://doi.org/10.1103/PhysRevLett.119.077202} {\path{doi:10.1103/PhysRevLett.119.077202}}.

\bibitem{Kurimune2020}
Y.~Kurimune, M.~Matsuo, Y.~Nozaki, Observation of gyromagnetic spin wave resonance in nife films, Phys. Rev. Lett. 124 (2020) 217205.
\newblock \href {https://doi.org/10.1103/PhysRevLett.124.217205} {\path{doi:10.1103/PhysRevLett.124.217205}}.

\bibitem{tateno2021Phys.Rev.B}
S.~Tateno, Y.~Kurimune, M.~Matsuo, K.~Yamanoi, Y.~Nozaki, {Einstein--de {{Haas}} Phase Shifts in Surface Acoustic Waves}, Phys. Rev. B 104 (2021) L020404.
\newblock \href {https://doi.org/10.1103/PhysRevB.104.L020404} {\path{doi:10.1103/PhysRevB.104.L020404}}.

\bibitem{Takahashi2020}
R.~Takahashi, H.~Chudo, M.~Matsuo, K.~Harii, Y.~Ohnuma, S.~Maekawa, E.~Saitoh, Giant spin hydrodynamic generation in laminar flow, Nat. Commun. 11 (2020) 3009.
\newblock \href {https://doi.org/10.1038/s41467-020-16753-0} {\path{doi:10.1038/s41467-020-16753-0}}.

\bibitem{Kazerooni2021}
H.~Tabaei~Kazerooni, G.~Zinchenko, J.~Schumacher, C.~Cierpka, Electrical voltage by electron spin-vorticity coupling in laminar ducts, Phys. Rev. Fluids 6 (2021) 043703.
\newblock \href {https://doi.org/10.1103/PhysRevFluids.6.043703} {\path{doi:10.1103/PhysRevFluids.6.043703}}.

\bibitem{Kazerooni2020}
H.~Tabaei~Kazerooni, A.~Thieme, J.~Schumacher, C.~Cierpka, {Electron Spin-Vorticity Coupling in Pipe Flows at Low and High Reynolds Number}, Phys. Rev. Appl. 14 (2020) 014002.
\newblock \href {https://doi.org/10.1103/PhysRevApplied.14.014002} {\path{doi:10.1103/PhysRevApplied.14.014002}}.

\bibitem{Zolfagharkhani-NatNano-2008}
G.~Zolfagharkhani, A.~Gaidarzhy, P.~Degiovanni, S.~Kettemann, P.~Fulde, P.~Mohanty, Nanomechanical detection of itinerant electron spin flip, Nat. Nanotechnol. 3 (2008) 720--723.
\newblock \href {https://doi.org/10.1038/nnano.2008.311} {\path{doi:10.1038/nnano.2008.311}}.

\bibitem{Funato2024}
T.~Funato, M.~Matsuo, T.~Kato, {Chirality-Induced Phonon-Spin Conversion at an Interface}, Phys. Rev. Lett. 132 (2024) 236201.
\newblock \href {https://doi.org/10.1103/PhysRevLett.132.236201} {\path{doi:10.1103/PhysRevLett.132.236201}}.

\bibitem{Callaway1959}
J.~Callaway, {Model for Lattice Thermal Conductivity at Low Temperatures}, Phys. Rev. 113 (1959) 1046--1051.
\newblock \href {https://doi.org/10.1103/PhysRev.113.1046} {\path{doi:10.1103/PhysRev.113.1046}}.

\bibitem{Ziman2001}
J.~M. Ziman, {Electrons and Phonons: The Theory of Transport Phenomena in Solids}, reissue Edition, Oxford University Press, 2001.

\bibitem{Ohe2024}
K.~Ohe, H.~Shishido, M.~Kato, S.~Utsumi, H.~Matsuura, Y.~Togawa, {Chirality-Induced Selectivity of Phonon Angular Momenta in Chiral Quartz Crystals}, Phys. Rev. Lett. 132 (2024) 056302.
\newblock \href {https://doi.org/10.1103/PhysRevLett.132.056302} {\path{doi:10.1103/PhysRevLett.132.056302}}.

\bibitem{Goehler2011}
B.~G{\"{o}}hler, V.~Hamelbeck, T.~Z. Markus, M.~Kettner, F.~Hanne, Z.~Vager, R.~Naaman, H.~Zacharias, {Spin Selectivity in Electron Transmission Through Self-Assembled Monolayers of Double-Stranded DNA}, Science 331 (2011) 894.
\newblock \href {https://doi.org/doi/10.1126/science.1199339} {\path{doi:doi/10.1126/science.1199339}}.

\bibitem{Naaman2012}
R.~Naaman, D.~H. Waldeck, {Chiral-Induced Spin Selectivity Effect}, J. Phys. Chem. 3 (2012) 2178--2187.
\newblock \href {https://doi.org/10.1021/jz300793y} {\path{doi:10.1021/jz300793y}}.

\bibitem{Michaeli2016}
K.~Michaeli, N.~Kantor-Uriel, R.~Naaman, D.~H. Waldeck, The electron{'}s spin and molecular chirality – how are they related and how do they affect life processes?, Chem. Soc. Rev. 45 (2016) 6478--6487.
\newblock \href {https://doi.org/10.1039/C6CS00369A} {\path{doi:10.1039/C6CS00369A}}.

\bibitem{Naaman2019}
R.~Naaman, Y.~Paltiel, D.~H. Waldeck, {Chiral molecules and the electron spin}, Nat. Rev. Chem. 3 (2019) 250--260.
\newblock \href {https://doi.org/10.1038/s41570-019-0087-1} {\path{doi:10.1038/s41570-019-0087-1}}.

\bibitem{Mishra2020}
S.~Mishra, A.~K. Mondal, S.~Pal, T.~K. Das, E.~Z.~B. Smolinsky, G.~Siligardi, R.~Naaman, {Length-Dependent Electron Spin Polarization in Oligopeptides and DNA}, J. Phys. Chem. C 124 (2020) 10776--10782.
\newblock \href {https://doi.org/10.1021/acs.jpcc.0c02291} {\path{doi:10.1021/acs.jpcc.0c02291}}.

\bibitem{Sano2024}
R.~Sano, T.~Kato, Chirality-induced spin selectivity by variable-range hopping along {DNA} double helix, arXiv:2404.19000 (2024).
\newblock \href {https://doi.org/10.48550/arXiv.2404.19000} {\path{doi:10.48550/arXiv.2404.19000}}.

\bibitem{Miwa2020}
S.~Miwa, K.~Kondou, S.~Sakamoto, A.~Nihonyanagi, F.~Araoka, Y.~Otani, D.~Miyajima, Chirality-induced effective magnetic field in a phthalocyanine molecule, Appl. Phys. Express 13 (2020) 113001.
\newblock \href {https://doi.org/10.35848/1882-0786/abbf67} {\path{doi:10.35848/1882-0786/abbf67}}.

\bibitem{Kondou2022}
K.~Kondou, M.~Shiga, S.~Sakamoto, H.~Inuzuka, A.~Nihonyanagi, F.~Araoka, M.~Kobayashi, S.~Miwa, D.~Miyajima, Y.~Otani, Chirality-induced magnetoresistance due to thermally driven spin polarization, J. Am. Chem. Soc. 144 (2022) 7302--7307.
\newblock \href {https://doi.org/10.1021/jacs.2c00496} {\path{doi:10.1021/jacs.2c00496}}.

\bibitem{Kondou2023}
K.~Kondou, S.~Miwa, D.~Miyajima, Spontaneous spin selectivity in chiral molecules at the interface, J. Magn. Magn. Mater. 585 (2023) 171157.
\newblock \href {https://doi.org/10.1016/j.jmmm.2023.171157} {\path{doi:10.1016/j.jmmm.2023.171157}}.

\bibitem{Miwa2024}
S.~Miwa, T.~Yamamoto, T.~Nagata, S.~Sakamoto, K.~Kimura, M.~Shiga, W.~Gao, H.~M. Yamamoto, K.~Inoue, T.~Takenobu, T.~Nozaki, T.~Ohto, Spin polarization driven by molecular vibrations leads to enantioselectivity in chiral molecules, arXiv:2412.03082 (2024).
\newblock \href {https://doi.org/10.48550/arXiv.2412.03082} {\path{doi:10.48550/arXiv.2412.03082}}.

\bibitem{Bruus2004}
H.~Bruus, F.~Flensberg, {Many-Body Quantum Theory in Condensed Matter Physics --- An Introduction}, {Oxford University Press}, 2004.

\end{thebibliography}
\end{document}